\documentclass[conference]{IEEEtran}

\usepackage{siunitx}

\IEEEoverridecommandlockouts
\usepackage{cite}
\usepackage{amsmath,amssymb,amsfonts}
\usepackage{algorithmic}
\usepackage{graphicx}
\usepackage{textcomp}
\usepackage{xcolor}
\usepackage{comment}
\usepackage{multirow}
\usepackage[caption=false]{subfig}
\usepackage{bm}
\usepackage{url}
\usepackage{xcolor}

\begin{document}
\title{The Effect of COVID-19 on the Transit System in Two Regions: Japan and USA 
\thanks{\textsuperscript{1} Ismail Arai was affiliated also as a visiting scholar at James Madison University, Virginia, USA, while working on parts of this paper.}}

\author{
\IEEEauthorblockN{Ismail Arai \textsuperscript{1}}
\IEEEauthorblockA{Information Initiative Center\\Nara Institute of Science and Technology\\ 
ismail@itc.naist.jp}
\and 
\IEEEauthorblockN{Samy El-Tawab, Ahmad Salman}
\IEEEauthorblockA{College of Integrated Science and Engineering \\
James Madison University \\
 \{eltawass,salmanaa\}@jmu.edu}
 \and
\IEEEauthorblockN{Ahmed Elnoshokaty}
\IEEEauthorblockA{
Computer Information Systems \\ 
Northern Michigan University \\
aelnosho@nmu.edu}
}

\maketitle

\begin{abstract}
The communication revolution that happened in the last ten years has increased the use of technology in the transportation world. Intelligent Transportation Systems wish to predict how many buses are needed in a transit system. With the pandemic effect that the world has faced since early 2020, it is essential to study the impact of the pandemic on the transit system. This paper proposes the leverage of Internet of Things (IoT) devices to predict the number of bus ridership before and during the pandemic. We compare the collected data from Kobe city, Hyogo, Japan, with data gathered from a college city in Virginia, USA. Our goal is to show the effect of the pandemic on ridership through the year 2020 in two different countries. The ultimate goal is to help transit system managers predict how many buses are needed if another pandemic hits.  
\end{abstract}


\IEEEpeerreviewmaketitle

\section{Introduction}
COVID-19 has hit the world, freezing the transportation life, significantly impacting many public transportation systems. With people returning slowly to normal life, authorities are trying to meet the transportation demands while avoiding overcrowds and under-utilization. Several researchers have studied the impact of COVID-19 on public transit demand~\cite{mogaji2020impact,el2020framework,liu2020impacts}. However, this demand differs from one state to another in the United States, let alone from one country to another. This is due to the fact that restrictive measures during the pandemic differ from one location to another depending on the number of cases in an area, preventive executive orders in various places, and even cultural considerations \cite{bian2021time}. 

In this paper, we study the effect of COVID-19 on the public transit system in two different areas. The selection of two different cultures emphasizes the effect of social responsibility as part of the equation for the ridership demand during a pandemic. With the enhancement in wireless technology, researchers can take advantage of reliable communication to provide an answer to the ridership/demand question pre-and during a pandemic~\cite{ahangari2020public,hu2021left}.

\begin{figure}[!t]
\centering
 \includegraphics[width=\linewidth]{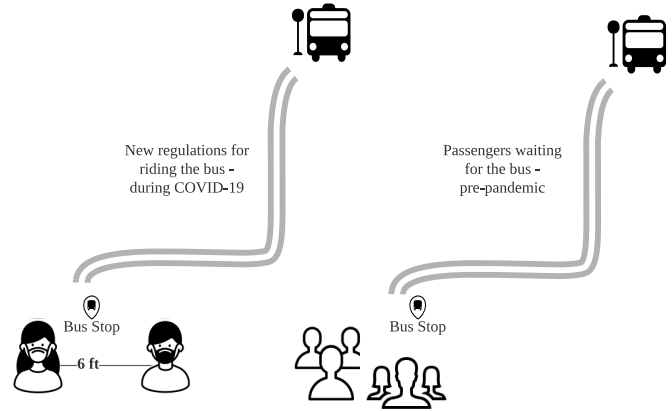}
 \caption{New Regulation for Bus Transit Systems during the Pandemic}
 \label{framework}
\end{figure}

We picked a college town (located in Harrisonburg, Virginia, USA) and a city in Japan (Kobe city in Hyogo, Japan) to be the locations for our study. We note that in Virginia, USA's commonwealth state, several stay-at-home orders were issued by the governor starting March 2020. To compare our data collection with regular ridership, we include in our study pre-pandemic ridership data. During the pandemic, several new regulations have been required to ride a public transit bus (e.g., wearing masks, social distance) as shown in Figure~\ref{framework}.   


 The remainder of the paper is organized as follows: section~\ref{sec:related} discusses what other researchers have done until now in estimating the public transit system demand in any city. Section~\ref{sec:ExperimentSetup} explains the data collection methods in the two locations (USA and Japan). In section~\ref{sec:results}, we discuss in detail the data analysis and results of our findings. Finally, we conclude the paper and shed light on future work directions in section~\ref{sec:final}.

\section{Related Work}\label{sec:related}
Several researchers have studied the idea of making a more sustainable urban transportation system. In the last decade, the wireless communication revolution allowed the integration of communication and public transportation to improve the quality of service (e.g., ridership and waiting time in public bus transit). Many researchers have also studied travel time and origin-destination estimation using Wi-Fi and Bluetooth~\cite{dunlap2016estimation, el2019origin, ryu2020wifi, oransirikul2016feasibility}. The use of Internet of Things (IoT) devices has increased to improve the quality of data estimate and ridership~\cite{el2017data,videa2019inference}. Other researchers have integrated the use of Machine Learning Techniques to analyze the relationship between data collected, and other factors (e.g., weather)~\cite{sousa2020iot,arai2021leveraging}. Previous studies demonstrated the accuracy of predicting ridership in subways and ride-sourcing through employing machine learning techniques~\cite{ding2016predicting,yan2020using}. 

With the recent hit of pandemic (COVID-19), the world of transportation has changed. Several preventive measures were taken to secure the use of the public transit system~\cite{kamga2021slowing}.  Wilbur et al.~\cite{wilbur2020impact} studied the effect of COVID-19 on ridership in two locations inside the United States (Nashville, Tennessee, and Chattanooga, Tennessee); in their work, they focused on comparing between highest-income areas and lowest-income areas. Palm et al.~\cite{palm2021riders} studied the social aspects of the COVID-19 on ridership. They aim to measure the burdens borne by riders who stopped using public transit during this pandemic period. Other researchers focused on the ridership analysis during the pandemic~\cite{ahangari2020public,leung2020data}. Ahangari et al.~\cite{ahangari2020public} investigated the effect in Baltimore and similar cities during the first five months of 2020. We believe that looking at data since March 2020 would help get a more accurate big picture of the situation and better results from the study.  

In this research, we study the effect of COVID-19 on ridership and compare the effect in two different countries. In addition, we leverage IoT devices to collect data in one of the two locations. The idea of two different countries is to highlight the cultural differences in the use of public transit systems during the pandemic.  

\section{System Architecture}\label{sec:ExperimentSetup}

The proposed system in our work utilizes IoT sensors that can collect data inside a bus on the number of passengers riding the bus. The collected data is used to build several applications for bus systems. Our case studies were located in Japan, and we compared our findings with another location in the USA. This paper investigates the correlation between COVID-19 cases and the data collected relative to buses’ ridership. 

In the USA, we chose James Madison University (JMU), with around $22,000$ on-campus students located in Harrisonburg, Virginia, that has a population of around $50,000$. The city depends on transit public bus systems as one of the main transportation aspects~\cite{el2020framework}. 

\begin{figure}[!ht]
 \begin{center}
  \includegraphics[width=\linewidth]{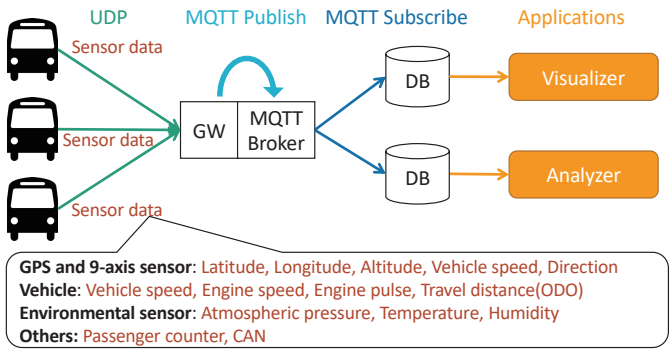}
  \caption{Data flow of the IoT Bus system in Japan.}
  \label{fig:framework_Japan}
 \end{center}
\end{figure}

In Japan, we collaborate with a bus company, Minato Kanko Bus Inc., which has developed its digital tachograph system named DOCOR~\cite{arai2020wfiot} to optimize its bus management efficiency. 
Figure~\ref{fig:framework_Japan} shows the data flow of DOCOR. Digital tachograph devices send sensor data to the Gateway (GW) with User Datagram Protocol (UDP) to avoid connection errors in an unstable mobile network. Once the GW receives the sensor data, all the data flow is via Transmission Control Protocol (TCP) based protocols of Message Queuing Telemetry Transport (MQTT) and Hypertext Transfer Protocol (HTTP). Every second, the sensor data has been stored in the Database (DB) since 2014. Additionally, the sampling interval was enhanced to $2$ Hz in $2018$. We have analyzed over $500$GB of the data for many purposes.

\begin{table}[t]
  \begin{center}
    \caption{Sensor data produced by DOCOR system.}
    \label{tab:sensors_docor}
    \begin{tabular}{ll}
      \hline
      Name & Source\\
      \hline
      Vehicle ID        &   License plate\\
      Route ID         &   Bus operation management system\\
      Vehicle ID of a route & Bus operation management system\\
      Driver ID     &   Bus operation management system\\
      Operation State & Manually input by drivers\\
      Date          &   Clock\\
      Time          &   Clock\\
      Latitude		&	GPS\\
      Longitude		&	GPS\\
      Altitude		&	GPS	\\
      Accuracy      &   GPS\\
      Travel distance (ODO)	&	Vehicle\\
      Travel distance (Trip) &  Vehicle\\
      Atmospheric pressure	&	Environmental sensor\\
      Temperature	&	Environmental sensor\\
      Humidity		&	Environmental sensor\\
      The number of boarding & Passenger counter\\
      The number of alighting & Passenger counter\\
      \hline
    \end{tabular}
  \end{center}
\end{table}

\begin{figure}[!t]
 \begin{center}
  \includegraphics[width=0.65\linewidth,clip]{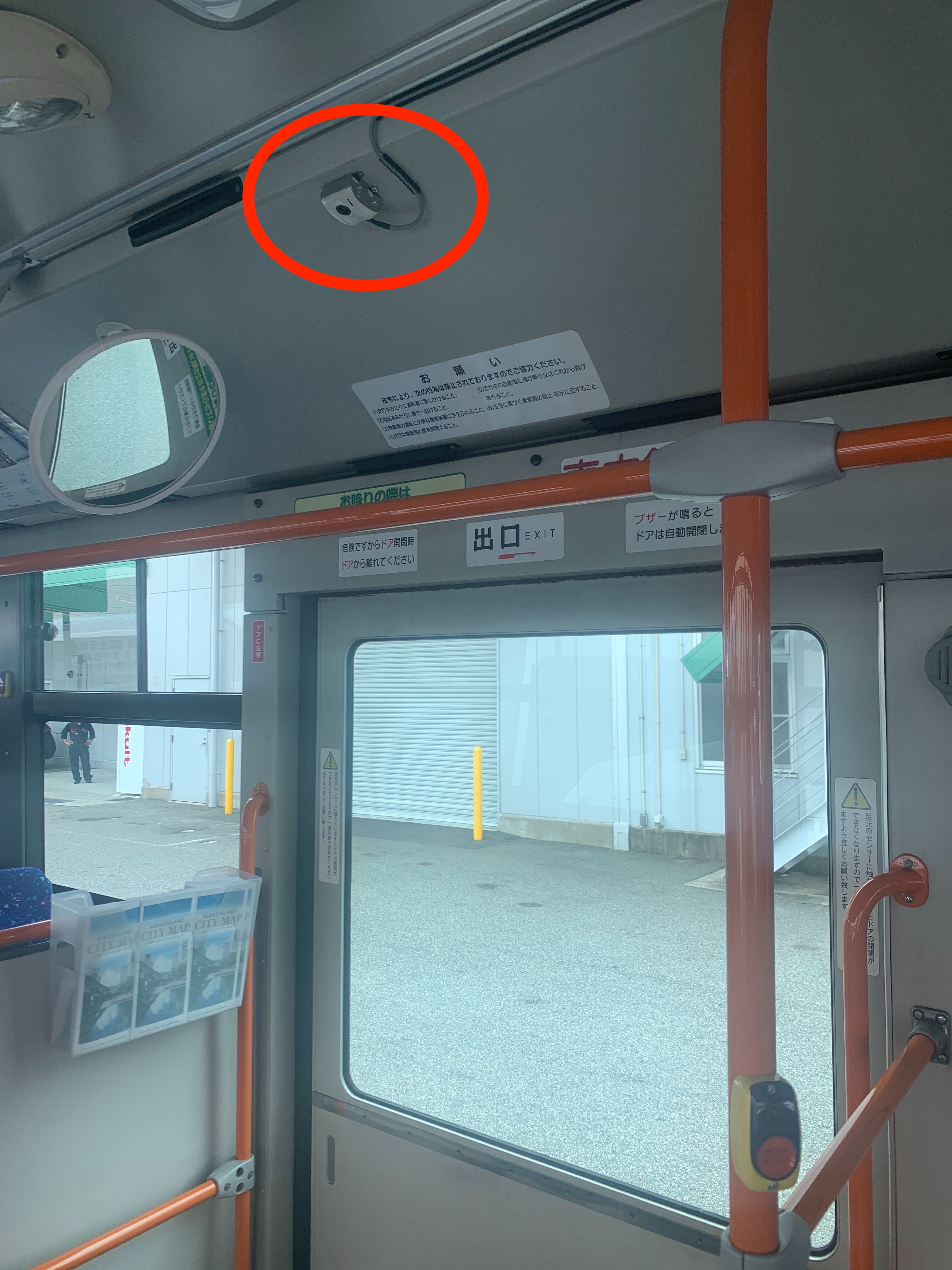}
  \caption{Passenger counter}
  \label{fig:passenger_counter}
 \end{center}
\end{figure}

\begin{figure*}[!ht]
 \subfloat[Ridership of a large bus]{
  \includegraphics[width=\textwidth]{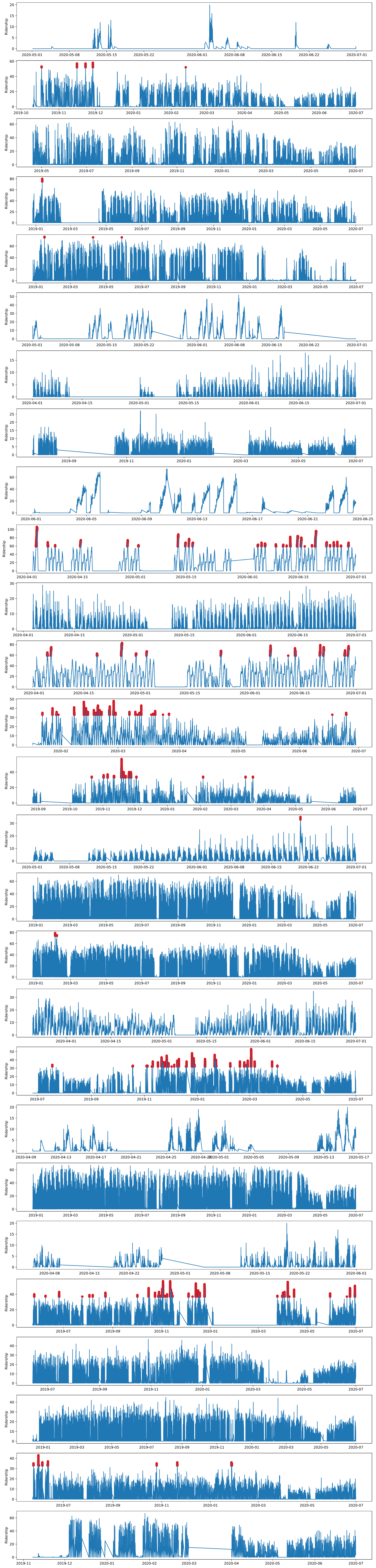}
 }
 
 \subfloat[Ridership of a small bus]{
  \includegraphics[width=\textwidth]{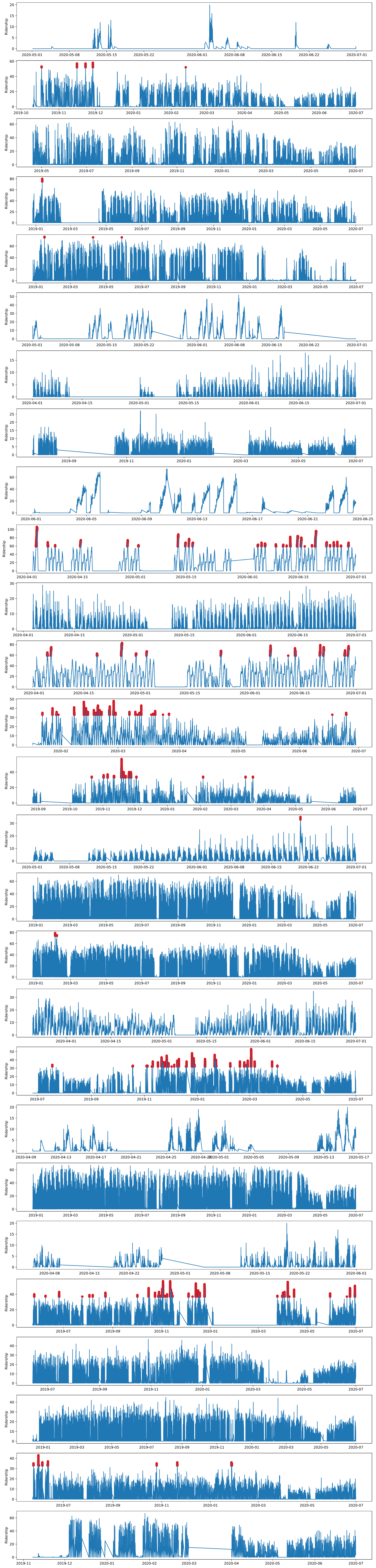}
 }
 \caption{Ridership of a Japanese company's buses.}
 \label{fig:ridership_graph_japan}
\end{figure*}

In this paper, we utilize the managed information on the system and sensor data, as shown in Table~\ref{tab:sensors_docor}. The essential sensor in this work is the designated passenger counter. As shown in Figure~\ref{fig:passenger_counter}, a camera is installed on the ceiling right above the door. The image processing unit inputs the camera image and outputs the number of boarding/alighting passengers. We confirmed it could count the ridership with around 98\% accuracy~\cite{NAF19}. Vehicle ID, route ID, date, time, and the environmental sensor data might correlate to ridership. The other sensor data in Table~\ref{tab:sensors_docor} are supplemental information to distinguish each route. In order to be able to understand the effect of COVID-19 on ridership, it is important to show the numbers in pre-and during the pandemic.

\begin{figure}[!ht]
 \begin{center}
  \includegraphics[width=\linewidth]{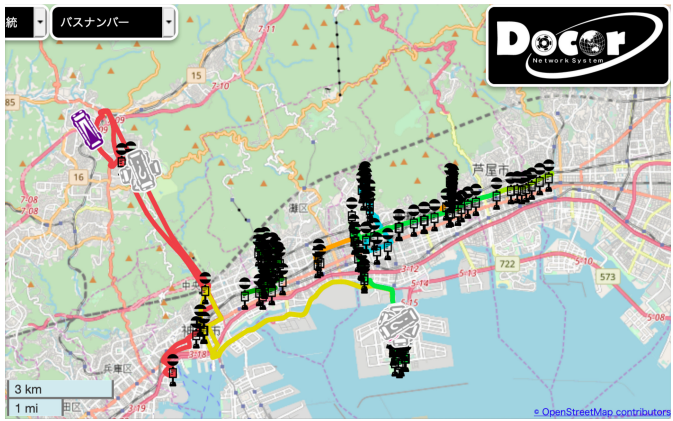}
  \caption{Bus routes in Japan.}
  \label{fig:bus_routes_japan}
 \end{center}
\end{figure}

Figure~\ref{fig:ridership_graph_japan} shows the ridership of two buses running in Kobe City, Japan. Red dots show the cases in which the passenger capacity is exceeded. A large bus is rarely packed, but a small one often is. The motivation in predicting ridership is to avoid that kind of packing. If a bus company can predict which buses will be crowded, they can run additional buses on those routes. 
The overview of the bus routes and the specifications are shown in Figure~\ref{fig:bus_routes_japan} and Table~\ref{tab:spec_bus}, respectively.

 \begin{table}[!ht]
   \begin{center}
     \caption{Specification of the route buses.}
     \label{tab:spec_bus}
     \begin{tabular}{lr}
       \hline
       Name & Value\\
       \hline
       Major routes & 8\\
       Minor routes & 40\\
       Bus stops & 106\\
       Buses & 77\\
       Buses with passenger counters & 27\\
       \hline
     \end{tabular}
   \end{center}
\end{table}


\begin{figure*}[!t]
 \includegraphics[width=\linewidth]{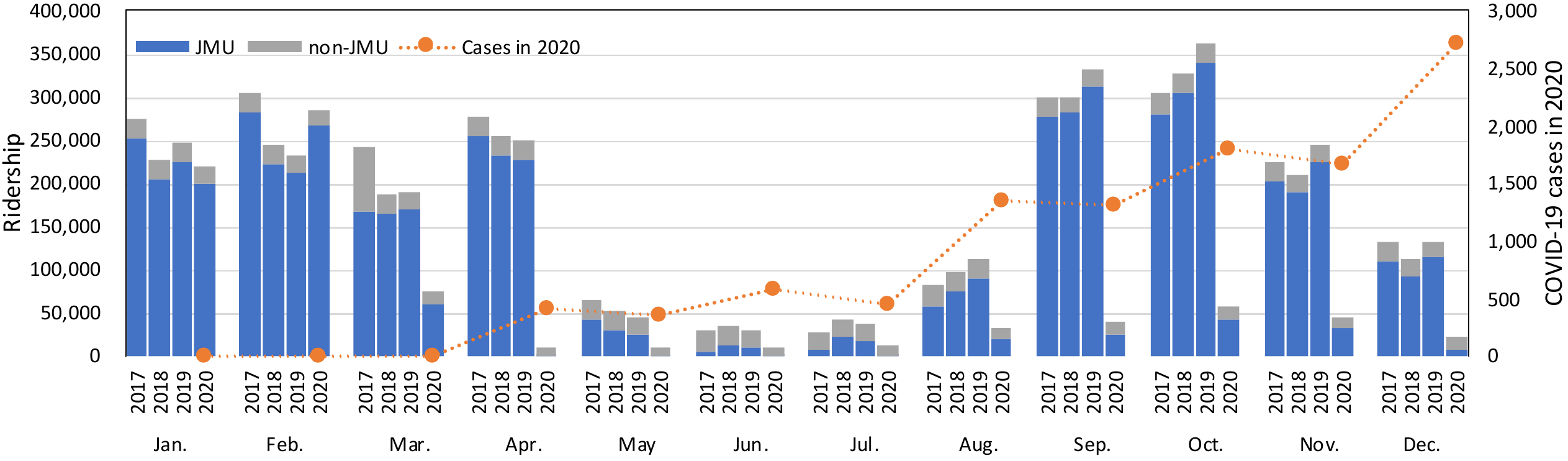}
 \caption{Impact of Monthly COVID-19 cases on ridership in Harrisonburg, USA.}
 \label{fig:covid19-vs-monthly-ridership-Harrisonburg}
\end{figure*}

\section{Results and Analysis} \label{sec:results} 
As mentioned in section~\ref{sec:related}, data collection in the USA was performed in the city of Harrisonburg, VA, USA, a mid-sized college town with a population of $50,000$. The bus transit system is one of the primary means of transportation in the city, specifically for the university. With the hit of the global pandemic in 2020, we highlight the effect of COVID-19 on the bus transit system. Figure~\ref{fig:covid19-vs-monthly-ridership-Harrisonburg} shows the number of ridership (in the city routes and James Madison University routes, respectively) for each month from 2017 to 2020. The ridership numbers for March, May, and June are always less than January, February, and April for a given year, except for 2020. The reason for that is that March includes one week of spring break when the vast majority of students leave campus and travel outside of the city. The same applies to May and June, when the regular academic year ends, and the summer break occurs. James Madison University students make up about 30\% of the Harrisonburg population, which is consistent with the decline in the percentage of ridership for the mentioned months. These numbers have also been consistent with data for the past ten years, according to the National Transit Database (NTD)~\cite{NTD}.

\subsection{Timeline and correlation in USA between COVID, Policies and Ridership}

The effects of the COVID-19 pandemic started in March 2020 for most territories in the USA. The city of Harrisonburg and the university were no exception in that stay-at-home and lockdown orders issued by the local government were put in effect.


Figure~\ref{fig:covid19-vs-monthly-ridership-Harrisonburg} shows the timeline of the critical policies and decisions that affected the public transit system in Virginia, USA, during COVID-19. As seen in Figure~\ref{fig:covid19-vs-monthly-ridership-Harrisonburg} on March 30, 2020, a stay-home order was issued, closing all schools and universities until the end of the academic year; this closing was kept until three phases of opening start happening on May $15^{th}$, June $5^{th}$ and July $1^{st}$, 2020 respectively. In phase $1$, people were urged to keep social distancing practice, work from home as possible, wear masks in public. Also, social activities were capped at $10$ people. Phase$2$ increased the number of social activities to $50$ people and allowed gyms to reopen at $30\%$ occupancy. Finally, phase $3$ increased social activities to $250$, allowing gyms to have $75\%$ occupancy. 

\begin{figure*}[!ht]
 \includegraphics[width=\linewidth,clip]{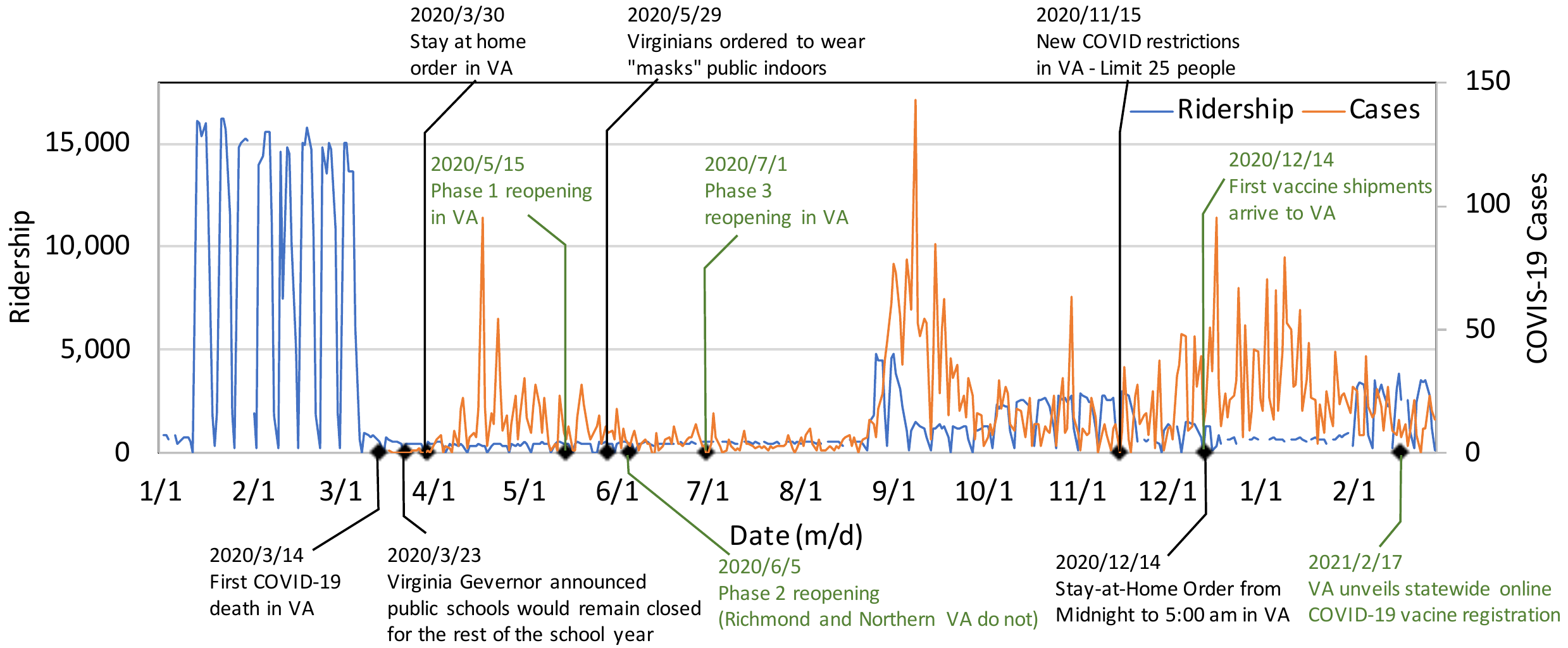}
 \caption{Impact of Daily COVID-19 cases and the related events on ridership in Harrisonburg, USA.}
 \label{fig:daily-covid19-vs-daily-ridership-Harrisonburg}
\end{figure*}

With the beginning of the fall 2020 semester, universities started on-campus. A spike of cases in early September forced an online switch for another month before switching to a hybrid-teaching mode.  It explains the massive drop in the number of ridership from March through June for the year 2020 for the city of Harrisonburg and the university, respectively, as shown in Figure~\ref{fig:covid19-vs-monthly-ridership-Harrisonburg}. 
 Figure~\ref{fig:daily-covid19-vs-daily-ridership-Harrisonburg} shows the number of cases reported in the city of Harrisonburg through the year of 2020. It is shown that the increase in case numbers correlates with a significant drop in the ridership numbers.    
 
\begin{figure*}[!ht]
 \includegraphics[width=\textwidth]{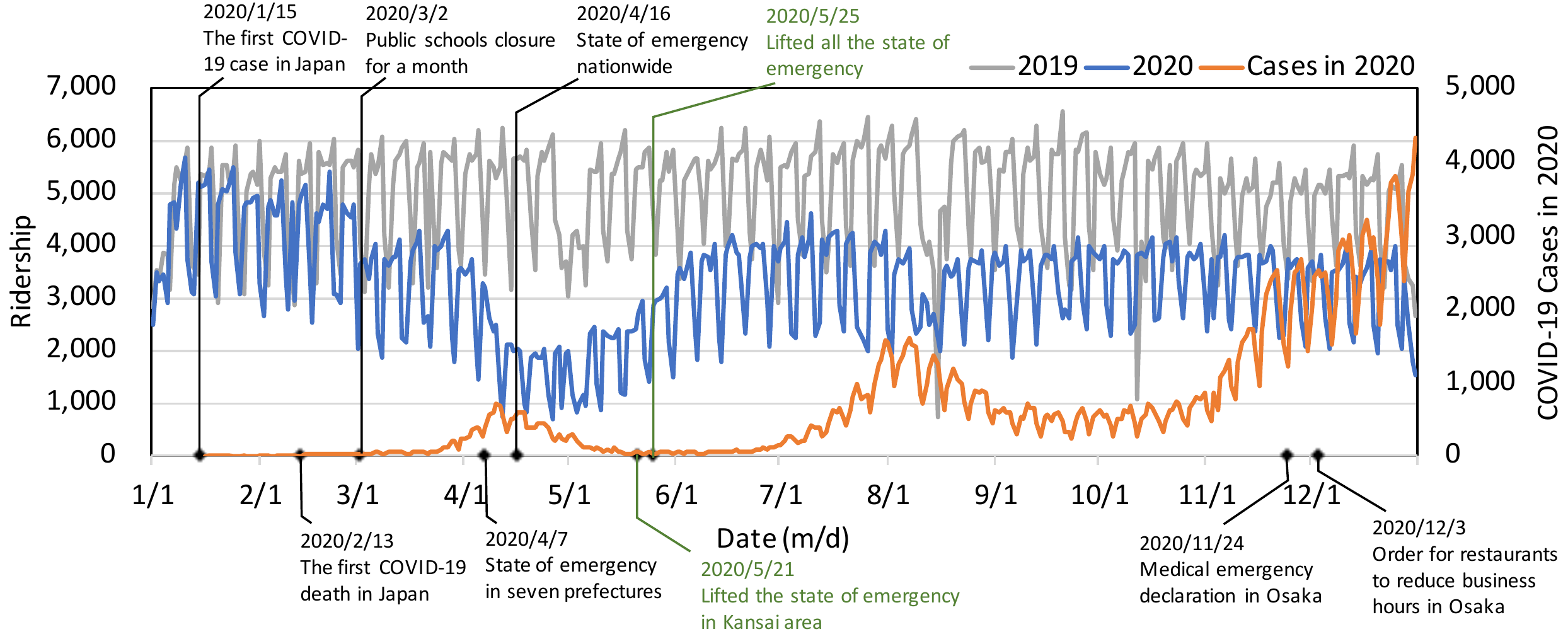}
  \caption{Impact of daily COVID-19 cases and the related events on ridership in Kobe, Japan.}
 \label{fig:covid19-vs-ridership-Japan}
\end{figure*}

\subsection{Timeline and correlation in Japan between COVID, Policies and Ridership}

Figure~\ref{fig:covid19-vs-ridership-Japan} simultaneously plots the daily ridership of Minato Kanko Bus in 2019 and 2020, the COVID-19 cases with related events in Japan in 2020. The ridership had remained stable between 3,000 and 6,000 in pre-pandemic 2019. This gap is due to commuters, so the ridership on weekdays is higher than on weekends. The ridership decreases in the Japanese long holidays in May, August, and the new year. The sudden decrease in mid-October was not a usual event, happening just in that year. After confirming the first COVID-19 case in Japan, the ridership started to decrease. The government's official first order of public schools closure affected students and their parents, dramatically dropping the ridership into two-thirds of a usual ridership of the last year. Japan faced the first impact of increasing cases at the end of March, then in a week, a state of emergency was issued for seven prefectures, including Hyogo prefecture. Even though the cases subsided in May, the ridership did not return to the pre-pandemic level. The ridership gradually increased, but after decreasing slightly when the cases started to increase in July, the ridership has been almost two-thirds of the pre-pandemic level. Every time the waves came, the ridership hardly changed anymore, despite the cases was increasing. People have learned to adapt to the pandemic by remote working and going out less to meet with people other than family members.

Our hypothesis of a negative correlation between ridership and COVID-19 cases was rejected. Even though Figure 8 shows three waves of increasing cases, only the first time was negatively correlated with the ridership. At that time, the government issued a state of emergency. It allows the prefectural governors to issue Stay-home-orders. They can also request or issue instructions to close schools or restrict the use of facilities that attract many people, such as department stores and movie theaters. While rapidly increasing cases at the end of November,  the government did not issue a state of emergency. So, Osaka prefecture, located next to Hyogo prefecture, had issued their original declaration. Despite that, the ridership did not decrease. Only the state of emergency issued by the government reduced the ridership in 2020. 

\subsection{Comparison between the two countries: Culture, Societal aspects, and Governmental Orders/Restrictions}
It is clear from the above analysis that the effect that happened on the public transit system (e.g., bus system) differs from one country to another. In the following, we highlight some of the cultural, societal, and governmental differences between (Virginia, USA) and (Kobe, Japan): 
\begin{itemize}
    \item \textbf{City differences:} Harrisonburg, Virginia, USA is a mid-size college-town city; the majority of the riders of the bus system are students, while Kobe, Japan, is a much bigger city with a population of $1.5$ million people. Despite the differences in size, we still believe that the comparison with historical data of each city shows enough evidence of the effect of COVID-19 on the public transit system in USA and Japan.
    \item \textbf{Cultures:} American society likes the use of their cars over public transit systems, while Japanese culture emphasizes the use of public transit systems as one of the environmental and social aspects.
    \item \textbf{Polices:} Bus systems in Harrisonburg, VA was on a reduced service - especially at the "Stay-home-orders" periods. Also, having online or hybrid teaching at James Madison University made most students return to their families home.  In contrast, the bus systems in Japan had a steady number of buses running during the pandemic.

    \item \textbf{Governmental Support:} In Virginia, due to the pandemic, the governmental unemployment benefits have been high enough to encourage many citizens to stay at home. Federal unemployment benefits authorized under the CARES Act ended September $4^{th}$, 2021 for Virginia residents. In Japan, $10,000$ yen ($935$ USD in June 2020) has only been given once to each citizen AFTER the first state of emergency.
\end{itemize}

\section{Conclusion and Future Work}\label{sec:final}
This paper studied the correlation between ridership and the level of pandemic severity. In our study, we used IoT devices installed in the bus in Japan as a method of collecting data about ridership during the pandemic period. To highlight that the pandemic has affected the transportation world differently, we compared our results with data collected from Harrisonburg, Virginia, USA. Our results show a clear correlation in both countries between COVID-19 cases and ridership. However, in Japan, people tend to return to using the public transit system faster than in the USA.  

Although our data analysis did not take into consideration the trend of the vaccination rate. We argue that any correlation of vaccination rates with ridership in our study could be neglected as the ridership data was collected up to early March 2021, where the vaccination rates were very low~\cite{cihan2021forecasting}. 

In future research, we plan to incorporate predictive modeling to accurately study the factors influencing the demand on bus ridership and compare our results with other cities inside the USA. In addition, as COVID-19 did not leave us yet, we wish to continue collecting data about the vaccination rate and how this affects the ridership population in both countries. Finally, our ultimate goal would be to estimate how many expected buses we need during a similar pandemic. Our vision is that autonomous-driven bus systems will be a future goal for transit systems~\cite{el2020autonomous}.



\section*{ACKNOWLEDGMENT}
This work was supported in part by the Jeffress Trust Awards Program in Interdisciplinary Research (Grant No. 550923) and by the Commonwealth Cyber Initiative (CCI), an investment in the advancement of cyber R\&D, innovation and workforce development (Grant no. E2235831). For more information about CCI, visit cyberinitiative.org; and JSPS KAKENHI Grant Numbers JP20K11789, JP20H04183. The authors want to thank Minato Kanko Bus Inc., Japan, for providing the sensor data and Mrs. Cheryl Spain from the City of Harrisonburg Public Transportation, USA, for providing the ridership data. 

\bibliographystyle{./bibliography/IEEEtran}
\bibliography{./bibliography/IEEEabrv,./bibliography/IEEEexample,./bibliography/reference.bib}

\end{document}